\documentclass[12pt]{article}

\usepackage[breaklinks]{hyperref}  
\usepackage{graphicx}
\usepackage{sariel,wide}
\usepackage{amstext}
\usepackage{amsmath}
\usepackage{enumerate}

\newcommand{\range}{\mathsf{R}}
\newcommand{\GroundSet}{\mathsf{X}}
\newcommand{\FGroundSet}{X}
\newcommand{\Set}{N}
\newcommand{\Family}{\EuScript{F}}

\newcommand{\VC}{\ensuremath{\mathsf{VC}}\xspace}
\newcommand{\VCDim}{\mathsf{d}}
\newcommand{\RangeSpace}{S}

\newcommand{\DistX}[2]{d_{#1}\pth{#2}}

\begin{document}
 
\title{Carnival of Samplings: Nets, Approximations, Relative and Sensitive}

\author{Sariel Har-Peled\SarielThanks{Work on this paper
      was partially supported by a NSF CAREER award
      CCR-0132901.}}

\date{\today}

\maketitle

\begin{abstract}
    We survey several results known on sampling in computational
    geometry.  
\end{abstract}


\section{Introduction}

In this write-up, we are interested in how much information can be
extracted by random sampling of a certain size for a range space of
\VC dimension $\VCDim$. In particular, we show that several standard
results about samplings follow from the sampling theorem of 
Lin \etal \cite{lls-ibscl-01}. 

The following assumes that the reader is familiar and comfortable with
$\eps$-nets and $\eps$-approximations. The results surveyed in this
write-up are summarized in \figref{results}.

\begin{figure}
    \begin{tabular}{|l|l|l|}
        \hline
        Name & Property $\forall \range \in \Family$ & Sample size \\
        \hline
        \hline
        \begin{tabular}{l}
            $\eps$-net 
            \cite{hw-ensrq-87}\\
            \thmref{epsilon:net}
        \end{tabular}
        &%
        $r(\range) \geq \eps \;\;\Rightarrow \;\; s(\range) >
        0$ & $\displaystyle 
        O\pth{\frac{d}{\eps} \log \frac{1}{\eps}}$ 
        \\
        \hline
        \begin{tabular}{l}
            $\eps$-approximation 
            \cite{vc-ucrfe-71}\\
            \thmref{epsilon:approximation}
        \end{tabular}
        & $
        \MakeVBig \cardin{r(\range) -s(\range) }
        \leq \eps$ %
        &%
        $\displaystyle O\pth{ \frac{\VCDim }{\eps^2}}$\\
        
        \hline
        \begin{tabular}{l}
            Sensitive
            $\eps$-approximation    \\
            \cite{b-dga-95, bcm-prsss-99}    \\
            \thmref{sensitive}
        \end{tabular}
        & 
        $\displaystyle \cardin{r(\range) - s(\range)}
        \leq \frac{\eps}{2} \pth{ \sqrt{r(\range)} + \eps}$
        &
        $\displaystyle O \pth{ \frac{\VCDim}{\eps^2} {
              \log \frac{1}{\eps} }}$
        \\
        \hline
        \begin{tabular}{l}
            Relative
            $(\eps,p)$-approximation    \\
            \cite{ckms-arers-06}\\
            \thmref{relative}
        \end{tabular}
        & 
        \hspace{-0.3cm}\begin{tabular}{l}
            $r(\range) \leq p \;\;\Rightarrow \;\; s(\range) \leq (1+\eps)p$\\
            $r(\range) \geq p \;\;\Rightarrow$\\
            $\;\;\;\;
            (1-\eps)
            r(\range) \le s(\range)\leq (1+\eps) r(\range)$
        \end{tabular}
        &
        $\displaystyle O \pth{
           \frac{\VCDim}{\eps^2p}{\log\frac{1 }{ p} }}$ \\
        \hline\hline
    \end{tabular}
    \caption{ Here, $\range$ denotes a range in the given range space,
       $r(\range)$ is the fractional weight of $\range$, and $s(\range)$
       is its fractional weight in the random sample. The samples have the
       required property (for all the ranges in the range space) with
       constant probability.}6

    \figlab{results}
\end{figure}

\section{Preliminaries}

Lin \etal \cite{lls-ibscl-01} consider more general functions, but in
the settings we are interested in, their result can be described as
follows. We are given a range space $\RangeSpace = \pth{\GroundSet,
   \Family}$ of \VC dimension $\VCDim$, where $\GroundSet$ is a point
set, and $\Family$ is a family of subsets of $\GroundSet$.  In out
settings, we will usually consider a finite subset $\FGroundSet
\subseteq \GroundSet$ and we will be interested in the range space
induced by $\RangeSpace$ on $\FGroundSet$. In particular, let $\Set$
be a sample of $\GroundSet$.  For a range $\range \in \Family$, let
\[
r = r(\range) = 
\frac{\cardin{\range \cap \FGroundSet
      }}{\cardin{\FGroundSet}} 
\;\;\text{ and }\;\;
s = s(\range ) = 
\frac{\cardin{\range \cap \Set 
      }}{\cardin{\Set}} .
\]
Intuitively, $r$ is the total weight of $\range$ in $\FGroundSet$,
while $s$ is the sample estimate for $r$.  For a parameter $\nu > 0$,
consider the distance function between real numbers
\[
\DistX{\nu}{ r, s} = 
\frac{\cardin{r-s}}{r+ s+ \nu}.
\]
\begin{theorem}[\cite{lls-ibscl-01}]
    Let $\alpha, \nu, \delta > 0$ be parameters, and let $\RangeSpace
    = \pth{\GroundSet, \Family}$ be a range space with \VC dimension
    $\VCDim$. Let $\FGroundSet \subseteq \GroundSet$ be a finite
    set. We have, that a random sample (with repetition) of size
    \[
    O \pth{\frac{1}{\alpha^2 \nu} \pth{ \VCDim \log \frac{1}{\nu} +
          \log \frac{1}{\delta}}}
    \]
    from $\FGroundSet$ has the property
    that
    \[
    \forall \range \in
    \Family \;\;\;\;
    \DistX{\nu}{r(\range), s(\range)} < \alpha.
    \]
    And this holds with probability $\geq 1-\delta$.

    \thmlab{sampler}
\end{theorem}

It is hard in the sea of parameters to see the trees, so let us play
with the parameters a bit.

\section{Getting the \TPDF{$\eps$}{epsilon}-net and
   \TPDF{$\eps$}{epsilon}-approximation theorems}

\begin{theorem}[\cite{hw-ensrq-87}, $\eps$-Net Theorem]
    A sample of size $O( (d/\eps) \log (1/\eps))$ from $\FGroundSet$,
    is an $\eps$-net of $\RangeSpace = \pth{\GroundSet, \Family}$,
    where $\FGroundSet \subseteq \GroundSet$, and this holds with
    constant probability.

    \thmlab{epsilon:net}
\end{theorem}

\begin{proof}
    Let $\alpha =1/4$, $\nu = \eps$, $\delta = 1/4$, and apply
    \thmref{sampler}. The sample size is
    \[
    O \pth{\frac{1}{\alpha^2 \nu} \pth{ \VCDim \log \frac{1}{\nu} +
          \log \frac{1}{\delta}}} = 
    O \pth{\frac{\VCDim}{\eps} \log \frac{1}{\eps}}.
    \]
    Now, let $\range \in \Family$ be a range such that $\cardin{\range
       \cap \FGroundSet } \geq \eps n$, where $n =
    \cardin{\FGroundSet}$, we have that 
    \[
    \DistX{\nu}{r(\range), s(\range)} < \alpha = 1/4
    \]
    (with constant probably for all ranges). Namely,
    \[
    \frac{\cardin{r(\range)-s(\range)}}{r(\range)+ s(\range)+ \eps} < 1/4.
    \]
    The bad case for us, here is that $r(\range) \geq \eps$, but 
    $s(\range) = 0$. But then, the above inequality becomes
    \[
    \frac{1}{2} = 
    \frac{\eps}{2\eps} \leq 
    \frac{\cardin{r(\range)}}{r(\range)+ \eps} < 1/4,
    \]
    which is, of course, false. Thus, it must be that $s(\range ) >
    0$, which implies that $\Set$ is indeed an $\eps$-net.
\end{proof}

\begin{theorem}[\cite{vc-ucrfe-71}, $\eps$-Approximation Theorem.]
    A sample of size 
    \[
    O\pth{ \frac{1}{\eps^2}\pth{\VCDim + \log \frac{1}{\delta}}}
    \]
    from $\FGroundSet$, is an $\eps$-approximation of $\RangeSpace =
    \pth{\GroundSet, \Family}$, where $\FGroundSet \subseteq
    \GroundSet$, and this holds with probability $\geq 1-\delta$.

    \thmlab{epsilon:approximation}
\end{theorem}
\begin{proof}
    Set $\alpha =\eps /4$ and $\nu =1/4$. We have, by
    \thmref{sampler}, that for any $\range \in \Family$, it holds
    \[
    \cardin{r - s} \leq \frac{\eps}{4} \pth{ r + s + \nu} \leq \eps,
    \]
    implying the claim.
\end{proof}

\section{Sensitive \TPDF{$\eps$}{epsilon}-approximation}

Another similar concept was introduced by \cite{bcm-prsss-99}.

\begin{defn}
    A sample $\Set \subseteq \FGroundSet$ is \emphi{sensitive
       $\eps$-approximation} if
    \[
    \forall \range \in \Family \;\;\; \cardin{r(\range) - s(\range)}
    \leq \frac{\eps}{2} \pth{ \sqrt{r(\range)} + \eps}.
    \]
\end{defn}

Observe that a set $\Set$ which is sensitive $\eps$-approximation is,
simultaneously, both an $\eps^2$-net and an $\eps$-approximation.

\medskip
The following theorem shows the existence of sensitive
$\eps$-approximation. Note that the bound on its size is (slightly)
better than the bound shown by \cite{b-dga-95, bcm-prsss-99}.
\begin{theorem}
    A sample $\Set$ from $\FGroundSet$ of size
    \[
    O \pth{ \frac{1}{\eps^2} \pth{
          \VCDim \log \frac{1}{\eps} + \log \frac{1}{\delta}}}.
    \]
    is a sensitive $\eps$-approximation, with probability $\geq
    1-\delta$.

    \thmlab{sensitive}
\end{theorem}

\begin{proof}
    Let $\nu_i = i \eps^2/800$, $\alpha_i = \sqrt{1/4i}$, for $i=1,
    \ldots, M = \ceil{800/\eps^2}$. As such, for $i=1,\ldots, M$, we
    have $ \alpha_i^2\nu_i = \eps^2 /1600$.  Consider a single random
    sample $\Set$ of size
    \[
    U = O \pth{ \frac{1}{\eps^2} \pth{ \VCDim \log \frac{1}{\eps} +
          \log \frac{M}{\delta}}} = O \pth{ \frac{1}{\eps^2} \pth{
          \VCDim \log \frac{1}{\eps} + \log \frac{1}{\delta}}}.
    \]
    It is a sample complying with \thmref{sampler}, with parameters
    $\nu_i$ and $\alpha_i$, with probability at least $1-\delta/M$,
    since
    \[
    O \pth{\frac{1}{\alpha_i^2 \nu_i} \pth{ \VCDim \log
          \frac{1}{\nu_i} + \log \frac{M}{\delta}}} = O
    \pth{\frac{1}{\eps^2} \pth{ \VCDim \log \frac{1}{\eps} + \log
          \frac{M}{\delta}}} = O(U).
    \]
    Namely, \thmref{sampler} holds for $\Set$, with probability at
    least $\delta$, for parameters $\alpha_i$ and $\nu_i$, for all
    $i=1,\ldots, M$.

    Next, consider a range $\range \in \Family$, such that $r =
    r(\range) \in [(i-1)\eps^2/800, i\eps^2/800]$ and $s = s(\range)$.
    We assume for the sake of simplicity of exposition that $i>1$, as
    this case can be handled similarly to the more general case. This
    implies that $ {\nu_i}/{2} \leq r \leq \nu_i$, and as such
    \begin{equation}
        \alpha_i r \leq \alpha_i \nu_i =
        \sqrt{ \alpha_i^2 \nu_i \nu_i}
        \leq \sqrt{ \alpha_i^2 \nu_i} \sqrt{2r }
        = \sqrt{\frac{\eps^2}{800}} \sqrt{2r} 
        \leq \frac{\eps\sqrt{r}}{20}.
        \eqlab{stupid}
    \end{equation}
    We have that
    \[
    \DistX{\nu_i}{ r, s} < \alpha_i \;\; \Rightarrow \;\;
    \frac{\cardin{r-s}}{r+ s+ \nu_i} < \alpha_i.
    \]
    If $s \leq \nu_i$, we have that
    \[
    \cardin{r-s} \leq 3 \nu_i \alpha_i\leq \frac{\eps \sqrt{r}}{3}
    \leq \frac{\eps}{2} \pth{ \sqrt{r} + \eps},
    \]
    which implies that $\Set$ is indeed sensitive
    $\eps$-approximation.  Otherwise, if $s \geq \nu_i \geq r$, then
    we have
    \begin{align*}
        s-r \leq \alpha_i\pth{ r +s +\nu_i} &\;\;\Rightarrow \;\;
        (1-\alpha_i)(s-r)-\alpha_ir \leq
        \alpha_i\pth{ r  +\nu_i} \\
        &\;\;\Rightarrow \;\; (1-\alpha_i)(s-r) \leq
        \alpha_i\pth{ 2r  +\nu_i} \\
        &\;\;\Rightarrow \;\; s-r \leq \frac{ \alpha_i\pth{ 2r +\nu_i}
        }{(1-\alpha_i)} \leq 2 \alpha_i\pth{ 2r +\nu_i},
    \end{align*}
    since $\alpha_i \leq 1/2$. As such, by \Eqref{stupid}, we have
    \[
    \cardin{s-r} \leq 6 \alpha_i \nu_i \leq 6 \frac{\eps\sqrt{r}}{20}
    \leq \frac{\eps\sqrt{r}}{2},
    \]
    which implies the claim.
\end{proof}

\medskip

Looking on the bounds of sensitive $\eps$-approximation as compared to
$\eps$-approximation, its natural to ask whether its size can be
improved, but observe that since such a sample is also an
$\eps^2$-net, and it is known that $\Omega(d/\eps^2 \log(1/\eps))$ is
a lower bound on the size of such a net \cite{kpw-atben-92}, this
implies that such improvement is impossible.

\section{Relative \TPDF{$\eps$}{epsilon}-approximation}

\begin{defn}
    A subset $\Set \subset \FGroundSet$ is a \emph{relative
       $(p,\eps)$-approximation} if for each $\range \in \Family$,
    we have:
    \begin{enumerate}[(i)]
        \item If $r(\range) \geq p$ then 
        \[
        \displaystyle (1-\eps)
        r(\range) \le s(\range)\leq (1+\eps) r(\range).
        \]

        \item If $r(\range) \leq p$ then $s(\range) \leq (1+\eps)p$.
    \end{enumerate}
\end{defn}

The concept was introduced by \cite{ckms-arers-06}, except that
property (ii) was not required. However, property (ii) is just an easy
(but useful) ``monotonicity'' property that holds for all the
constructions I am aware of.

There are relative approximations of size (roughly) $1/(\eps^2 p)$. As
such, relative approximation is interesting, in the case where $p <<
\eps$. Then, we can approximate ranges of weight larger than $p$ with
a sample that has only linear dependency on $1/p$. Otherwise, we would
have to use the regular $p$-approximations, and there the required
sample is of size (roughly) $1/p^2$.

\begin{theorem}
    A sample $\Set$ of size $\displaystyle O \pth{
       \frac{1}{\eps^2p}\pth{\VCDim \log\frac{1 }{ p} +
          \log\frac{1}{\delta}}}$ is a relative
    $(p,\eps)$-approximation with probability $\geq 1-\delta$.    

    \thmlab{relative}
\end{theorem}

\begin{proof}
    Set $\nu = p/2$, and $\alpha =\eps/9$, and apply 
    \thmref{sampler}. We get that, for any range $\range \in \Family$,
    such that $r = r( \range) \geq p$ and $s = s(\range)$, it holds
    \[
    \DistX{\nu}{r, s} = \frac{\cardin{r-s}}{r+s+\nu} < \alpha
    \;\; \Rightarrow    \;\;%
    \cardin{r-s} \leq 
    \frac{\eps}{9} (r + s + p/2).
    \]
    If $s \leq r$ then, since $r \geq p$, we have that
    \[
    \cardin{r-s} \leq 
    \frac{\eps}{9} (r + s + p/2)
    \leq 
    \frac{\eps}{9} 3r
    \leq\eps r,
    \]
    which implies property (i). Otherwise, $s \geq r$, and then
    \[
    s - r \leq 
    \frac{\eps}{9} (r + s + p/2)
    \;\;\Rightarrow \;\; 
    (1-\eps/9)s  \leq (1+\eps/9) r + p/2.
    \]
    This implies that 
    \begin{equation}
        s \leq \frac{1+\eps/9}{1-\eps/9} r + \frac{1}{2 (1-\eps/9)}p
        \leq \frac{4}{3} r + \frac{9}{16} p
        \leq 2 r.
        \eqlab{silly:2}
    \end{equation}
    Thus, $s \leq 3r$, which implies that
    \[
    \cardin{r-s} \leq 
    \frac{\eps}{9} (r + s + p/2)
    \leq     \frac{\eps}{9}  5r 
    \leq 
    \eps r,
    \]
    which again implies  property (i).    

    As for property (ii), if $r \leq p$ then we need to show that $s
    \leq (1+\eps)p$, and this follows easily from the above
    calculations.
\end{proof}

In fact, one can slightly strengthen the concept by making
it ``sensitive''.  
\begin{theorem}
    A sample $\Set$ of size $\displaystyle O \pth{
       \frac{1}{\eps^2p}\pth{\VCDim \log\frac{1 }{ p} +
          \log\frac{1}{\delta}}}$ is a relative
    $(i p,\eps/\sqrt{i})$-approximation with probability $\geq
    1-\delta$, for all $i \geq 0$.    

    Namely, for any range $\range \in \Family$, such that $r(\range)
    \geq i p$, we have
    \begin{equation} 
        \pth{1-\frac{\eps}{\sqrt{i}}}
        r(\range)  \le
        s(\range)\leq
        \pth{1 + \frac{\eps}{\sqrt{i}}} r(\range).
        \eqlab{rel-app:2}
    \end{equation}
    
    \thmlab{relative:sensitive}
\end{theorem}
\begin{proof} 
    Set $p_i = i p$ and $\eps_i = \eps/\sqrt{i}$, for
    $i=1,\ldots,1/p$. Now, apply \thmref{relative}, and observe that all
    the samples are of the same size, and as such one can use the same
    sample to get this guarantee for all $i$.
\end{proof}

\bigskip

Interestingly, sensitive approximation imply relative approximations.
\begin{lemma}
    Let $\eps, p>0$ be parameters, and let $\eps' = \eps
    \sqrt{p}$. Then, if $\Set$ is sensitive $\eps'$-approximation
    to the set system $\pth{\FGroundSet, \Family}$ then its 
    also a relative $(\eps,p)$-approximation.
\end{lemma}

\begin{proof}
    We know that $\forall \range \in \Family$ it holds $\displaystyle
    \cardin{r(\range) - s(\range)} \leq \frac{\eps'}{2} \pth{
       \sqrt{r(\range)} + \eps'}$. As such, for $\range \in \Family$,
    if $r(\range) = \alpha p$ and $\alpha \geq 1$, then we have
    \[
    \cardin{r(\range) - s(\range)} \leq \frac{\eps\sqrt{p}}{2} \pth{
       \sqrt{\alpha p} + \eps\sqrt{p}}
    = \frac{\eps^2p}{2} +  \frac{\eps}{2} \sqrt{\alpha} p 
    \leq \pth{\frac{\eps^2}{2} + \frac{\eps}{2}} \alpha p \leq \eps
    r(\range),
    \]
    which implies that $\Set$ is a relative $(\eps,p)$-approximation.
\end{proof}

\medskip



 
\bibliographystyle{alpha} 
\bibliography{shortcuts,geometry}

\end{document}